# An Extreme Learning Machine-Based System Frequency Nadir Constraint Linearization Method

Likai Liu, *Student Member, IEEE*, Zechun Hu, *Senior Member, IEEE*, Nikhil Pathak, *Member, IEEE*, and Haocheng Luo, *Student Member, IEEE*

*Abstract*—Large-scale integration of converter-based renewable energy sources (RESs) into the power system will lead to a higher risk of frequency nadir limit violation and even frequency instability after the large power disturbance. Therefore, it is essential to consider the frequency nadir constraint (FNC) in power system scheduling. Nevertheless, the FNC is highly nonlinear and nonconvex. The state-of-the-art method to simplify the constraint is to construct a low-order frequency response model at first, and then linearize the frequency nadir equation. In this letter, an extreme learning machine-based network is built to derive the linear formulation of FNC, where the two-step fitting process is integrated into one training process and more details about the physical model of the generator are considered to reduce the fitting error. Simulation results show the superiority of the proposed method on the fitting accuracy.

## NOMENCLATURE

| | |
|---|---|
| $i, \mathcal{I}$ | Index and set of traditional generators (TGs). |
| $j, \mathcal{J}$ | Index and set of renewable energy resources (RESs). |
| $H, D$ | System inertia level and load damping factor. |
| $T_i^{r/g/c}$ | Reheater/governor/turbine time constants of TG $i$. |
| $F_i^g$ | High-pressure turbine fraction of TG $i$. |
| $T_j^v$ | Converter time constant of RES $j$. |
| $R_{i/j}^{g/v}$ | Differential/droop coefficient of TG $i$/RES $j$. |
| $x_i^g, x_j^v$ | Binary variables indicating the on-off status of TG and whether RES participating frequency control |

## I. INTRODUCTION

THE increasing penetration of converter-based renewable energy sources (RESs) has led to reduced system inertia, making it difficult to satisfy the frequency nadir limit after the large power imbalance [1], e.g., equipment failure. Therefore, the frequency nadir limit has attracted concerns in power system scheduling problems, such as unit commitment (UC), optimal power flow, economical dispatch, etc. Nevertheless, the closed-form expression of frequency nadir constraint (FNC) is highly nonlinear and nonconvex [2], making the power system scheduling model including the FNC become a nonlinear programming problem which is hard to solve.

To preserve the linearity of the power system scheduling models, several methods have been proposed to simplify the FNC. The first kind of method assumes that the regulation resource changes its output with a constant ramp rate during the primary frequency control process [3], [4], which is not consistent with the droop control method adopted by the frequency regulation resources. For the second kind of method [5]–[7], there are usually two steps in the derivation of FNC. The first step is to approximate the full order system frequency response model by a low order model, and the second step is to fit the expression of frequency nadir by utilizing the piecewise linearization technique. However, the fitting error may be superimposed and amplified in the two-step fitting process.

The authors are with the Department of Electrical Engineering, Tsinghua University, Beijing, 100084, China (e-mail: zechhu@tsinghua.edu.cn).

In this letter, a meticulously designed extreme learning machine (ELM)-based network is used to derive the FNC. The innovations of this method lie in two aspects: first, the two coordinated fitting processes in the two-step fitting method are integrated under the machine learning framework, which helps to avoid the superposition and amplification of the fitting error; second, the governor and turbine models of the traditional generator (TG) usually omitted by the existing methods are considered in the proposed method, and these models have notable influences on the frequency nadir according to [2].

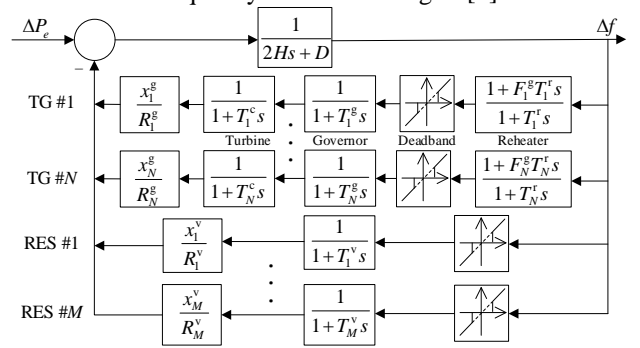

Fig. 1 Multi-machine centralized system frequency response model.

## II. FREQUENCY NADIR CONSTRAINT LINEARIZATION METHOD

### A. Frequency Response Model and Frequency Nadir Constraint

A centralized frequency response system comprised of TGs and converter-based RESs is depicted in Fig. 1. To derive the closed-form expression of frequency nadir, the authors in [7], [8] simplified the original model to a second-order frequency response system by omitting the deadband of the frequency controller, the governor and turbine blocks of TG, and assuming that $T^r = T \gg T^v \approx 0$. The transfer function of the simplified second-order frequency response system is as follows:

$$G(s) = \frac{\Delta f}{\Delta P_e} = \frac{1}{2HT} \frac{1+sT}{s^2 + 2\zeta\omega_n s + \omega_n^2} \quad (1)$$

where $\zeta$ and $\omega_n$ are the damping ratio and natural frequency of the second-order system.

Based on the transfer function model, the mathematic expression of FNC after a stepwise disturbance $\Delta P_e(s) = -\Delta P / s$ is given by (2).

$$\Delta f^{\text{nadir}} = -\frac{\Delta P}{D+R}\left(1+\sqrt{\frac{T(R-F)}{2H}}e^{-\zeta\omega_n t_m}\right) \leq \Delta f_{\max}^{\text{nadir}}, \quad (2)$$

where $R$ is the integrated droop coefficient of all online frequency regulation resources, $F$ is the integrated high-pressure turbine fraction of online TGs, $t_m$ is the time instance of the frequency nadir. The derivations of (1) and (2) are omitted due to space limitations, and the readers can refer to [7], [8] for the details.

All the variables in the middle part of (2) except for the time constant $T$ are functions of the binary variables $x_i^g$ and $x_j^v$, which are both decision variables in the power system optimi-zation model. Thus, the above FNC is highly nonlinear, making the scheduling model containing this constraint hard to solve.

### B. System Frequency Security Margin Forecast Model

It can be found from (2) that the frequency nadir is in direct proportion to the power disturbance. Therefore, the frequency deviation threshold $\Delta f_{\max}^{\text{nadir}}$ corresponds to a maximum tolerable power disturbance $\overline{\Delta P}$, which is termed as frequency security margin [7]. The frequency security margin denotes the maximum power disturbance the system can withstand without violating the frequency nadir limit. Based on this, FNC can be established by limiting the power outputs of generators below the frequency security margin.

Despite the deep learning methodology has developed rapidly in recent years [9], the deep learning algorithms cannot get a linear approximation of the FNC. Thus, an ELM-based forecast model is proposed to predict the frequency security margin in this study. The structure of the network is illustrated in Fig. 2, given as follows.

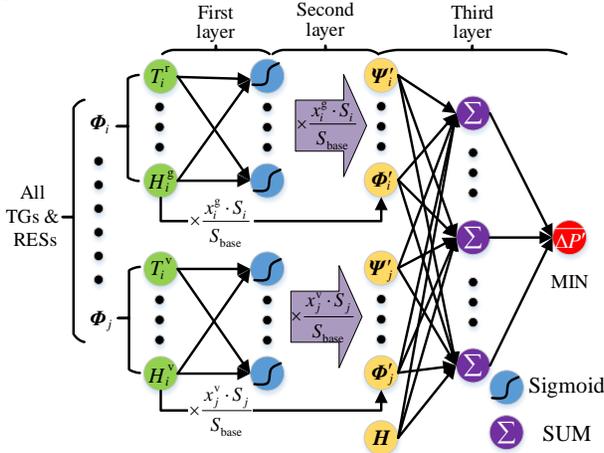

Fig. 2 ELM-based forecast model for the frequency security margin.

In the first layer, the original parameters of each TG/RES are inputted to a fully-connected layer using the sigmoid activation function $\vartheta$. The connection weight matrixes $\boldsymbol{a}^{\text{g/v}}$ and the bias vectors $\boldsymbol{b}^{\text{g/v}}$ for TG/RES are randomly generated. The outputs of the first layer are used as the new characteristic parameters of TG/RES, as

$$\boldsymbol{\Psi}_{i/j} = \vartheta\left(\boldsymbol{a}^{\text{g/v}} \cdot \boldsymbol{\Phi}_{i/j} + \boldsymbol{b}^{\text{g/v}}\right), \quad \forall i \in \mathcal{I}, j \in \mathcal{J}, \quad (3)$$

where $\boldsymbol{\Phi}_i \triangleq \{T_i^r, T_i^g, T_i^c, F_i^g, R_i^g, H_i^g\}$, $\boldsymbol{\Phi}_j \triangleq \{T_j^v, R_j^v, H_j^v\}$.

In the second layer, the new characteristic parameters along with the original parameters are multiplied by the decision variables $x_i^g$ and $x_j^v$ indicating the on-off status of TG and whether RES participating frequency control, as

$$\boldsymbol{\Psi}_i' = x_i^g \cdot \frac{S_i}{S_{\text{base}}} \cdot \boldsymbol{\Psi}_i, \quad \boldsymbol{\Phi}_i' = x_i^g \cdot \frac{S_i}{S_{\text{base}}} \cdot \boldsymbol{\Phi}_i, \forall i \in \mathcal{I}, \quad \text{(4-a)}$$

$$\boldsymbol{\Psi}_j' = x_j^v \cdot \frac{P_j}{S_{\text{base}}} \cdot \boldsymbol{\Psi}_j, \quad \boldsymbol{\Phi}_j' = x_j^v \cdot \frac{P_j}{S_{\text{base}}} \cdot \boldsymbol{\Phi}_j, \forall j \in \mathcal{J}, \quad \text{(4-b)}$$

where $S_i$ is the capacity of TG $i$, $P_j$ is the power generation of TG $j$, $S_{\text{base}}$ is the base power.

The total system inertia is calculated as (5) in the second layer of the forecast model and used as part of the input of the third layer. Because other electric devices like electromotor and synchronous condenser also provide inertia, and the total system inertia has important effects on the frequency nadir.

$$H = \frac{1}{S_{\text{base}}} \cdot \left[\sum_{i \in \mathcal{I}}\left(H_i^g \cdot S_i \cdot x_i\right) + \sum_{j \in \mathcal{J}}\left(H_j^v \cdot P_j \cdot x_j\right) + \sum_{e \in \mathcal{E}}\left(S_e \cdot H_e\right)\right], \quad (5)$$

In the third layer, the piecewise linearization technique is utilized to fit the frequency security margin, as

$$\overline{\Delta P'} = \min_{1 \leq l \leq \mathcal{L}}\left(\boldsymbol{c}_l \cdot [\boldsymbol{\Psi}', \boldsymbol{\Phi}', H]^{\top} + h_l\right), \quad (6)$$

where $l$ and $\mathcal{L}$ are the index and number of segments of the piecewise linear function, respectively, $\boldsymbol{c}_l / h_l$ is the coefficient vector/scalar of the first-degree/constant item of the piecewise linear function, $\boldsymbol{\Psi}'$ is a vector composed of every TG's $\boldsymbol{\Psi}_i'$ and every RES's $\boldsymbol{\Psi}_j'$, $\boldsymbol{\Phi}'$ is similar to $\boldsymbol{\Psi}'$.

The proposed ELM-based forecast model is trained by solving the optimization problem below.

$$\begin{aligned}\min_{\boldsymbol{c},\boldsymbol{h}} \quad & \sum_{k=1}^{\mathcal{K}}\left\{\min_{1 \leq l \leq \mathcal{L}}\left(\boldsymbol{c}_l \cdot [\boldsymbol{\Psi}', \boldsymbol{\Phi}', H]_k^{\top} + h_l\right) - \overline{\Delta P_k}\right\} \\ \text{s.t.} \quad & \min_{1 \leq l \leq \mathcal{L}}\left(\boldsymbol{c}_l \cdot [\boldsymbol{\Psi}', \boldsymbol{\Phi}', H]_k^{\top} + h_l\right) - \overline{\Delta P_k} \leq 0, \quad \forall 1 \leq k \leq \mathcal{K},\end{aligned} \quad (7)$$

where $k$ and $\mathcal{K}$ are the index and the number of training samples, $\overline{\Delta P_k}$ is the frequency security margin of the $k^{\text{th}}$ data sample. The constraint is added because a positive prediction error of the frequency security margin may result in the frequency nadir limit violation.

The training of the forecast model is offline and only need to be completed once. Then, the forecast model can be applied to develop the FNC and applied in the operation optimization.

*Remark 1*: By assuming that the damping coefficient $D$ is in direct proportion to the load power, and choosing the load power as the power base value $S_{\text{base}}$, the normalized damping coefficient is a constant and can be omitted from the model.

*Remark 2*: Although the proposed method does not explicitly consider the effects of governor deadband, it can learn the effects from the training data which contains the influences of the governor deadband on the frequency nadir.

### C. Formulation of the ELM-based Frequency Nadir Constraint

In the power system operation, the largest power disturbance is the tripping of the generator. The system frequency security margin after the tripping of the TG $i$ (denoted as $\overline{\Delta P_{\backslash i}'}$) should be larger than the scheduled power generation of this TG to ensure the FNC, as

$$\overline{\Delta P_{\backslash i}'} = \min_{1 \leq l \leq \mathcal{L}}\left(\boldsymbol{c}_l \cdot [\boldsymbol{\Psi}_{\backslash i}', \boldsymbol{\Phi}_{\backslash i}', H_{\backslash i}]^{\top} + h_l\right) \geq P_i, \quad \forall i \in \mathcal{G}. \quad (8)$$

where subscript $\backslash i$ denotes unit $i$ is offline, and $P_i$ in the right side of constraint is the power output of TG $i$. This constraint should be satisfied for every generator to guarantee that the FNC will not be violated after the failure of any single generator.

The obtained FNC is a piecewise linear affine function of the decision variables $x_i^g$ and $x_j^v$ in the power system scheduling model.

## III. CASE STUDY

### A. Case Settings

The IEEE 118-bus system is used as the test system. We modified it by connecting three wind farms with installed capacities of 400 MW, 300 MW, 300MW, and one photovoltaic power station with an installed capacity of 330 MW. Generator parameters used in this research are taken from [10] and [11].

Considering that FNC is usually used in the UC problem, we simulated the UC over a whole year on the test system by utilizing the model proposed in subsection 4.2.3 of [12]. Historical data of load power and renewable generations for one year from a real power system in northern China are scaled down to fit the test system and then used as the input for the UC simulation. A REG will neither participate primary frequency control nor provide inertia when its power output is lower than 0.3 pu.

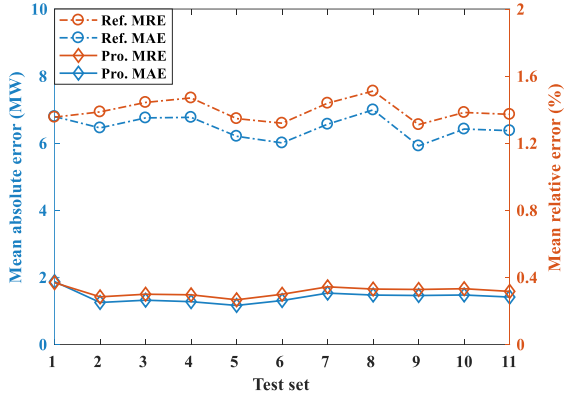

Fig. 3 Fitting errors of the two methods obtained from different test sets.

TABLE I Average performances of different methods on all test sets.

| Method | Average MAE | Average MRE | Average Training Time |
| --- | --- | --- | --- |
| Proposed | 1.4132 MW | 0.31 % | 323.43 s |
| Reference [13] | 6.4788 MW | 1.39 % | 3.84 s |

After the UC simulation, a total of 8760 data samples are obtained, and each data sample is composed of binary variables indicating the on-off status of TG and whether RES participating frequency control. To make a rigorous statistical evaluation of the performance of the proposed method, some noises are added to the data set obtained from the UC simulation to generate another ten data sets. In order to eliminate the influences of time and season, for each data set, 8000 data samples are *randomly* chosen for training and the rest 760 data samples are used for testing.

The detailed system frequency response model shown in Fig. 1 is used to simulate the frequency security margin of the test system for each data sample. This frequency response model contains the governor deadband, so the influence of the governor deadband to the proposed method can be tested.

*B. Simulation Result and Analysis*

Since the two-step fitting method has shown its superiority in accuracy compared with its previous methods in [13], the proposed method is only compared with the two-step fitting method in this letter. Some settings about these two methods are shown as follows.

*Proposed method:* The neuron number in the hidden layer of the ELM is 10 for each TG/RES. The number of segments of the piecewise linear function is 3, i.e. $\mathcal{L} = 3$ in (6).

*Two-step fitting method:* The number of segments for the piecewise linear function is 40.

The performance stability of the ELM-based method may be questioned because the connection weight matrixes and bias vectors are randomly generated. To clearfy this issue, the proposed method is trained and tested ten times for each data set, and the performances of the proposed method given hereinafter are the average results of ten tests.

The comparisons of the two methods are from two aspects: fitting accuracy and training time. The mean absolute errors (MAEs) and mean relative errors (MREs) of different methods are shown in Fig. 3. Table I gives the average MAE, MRE, and training time on all test sets.

It is observed in Fig. 3 that for all data sets, the fitting errors of the proposed method are significantly smaller than those of the two-step fitting method. Table I also shows that both the average MAE and MRE of the proposed method are smaller than these of the two-step fitting method. The fluctuation of the fitting error of the proposed method is quite small, which proves that the performance of the proposed method is stable.

The average training time of the proposed method is larger than that of the two-step fitting method, but the longer training time can be overlooked in view of the accuracy merits because the training process is offline.

Besides, these test results illustrate that the proposed method is applied for the frequency response model containing governor deadzone.

IV. CONCLUSION

Large-scale integration of converter-based RES has increased the risk of frequency nadir limit violation and frequency instability, so the FNC should be incorporated into the power system scheduling problem. To linearize the formulation of FNC, this letter has proposed an ELM-based network to forecast the frequency security margin of the power system. Case studies show that, compared with the two-step fitting method, the fitting accuracy of the proposed method is significantly higher, which proves that the proposed method provides a closer approximation to the original FNC.

The inertia distribution in a synchronous grid may be inhomogeneous, which may challenge the applicability of centralized frequency response model and affect the accuracy of the proposed method. In the future work, this method will be developed for the grid with inhomogeneous inertia distribution.